\input harvmac
\input epsf

\noblackbox 

\Title
 {\vbox{
 \baselineskip12pt
 \hbox{HUTP-01/A056}
 \hbox{hep-th/0111051}\hbox{}\hbox{}
}}
{\vbox{
 \centerline{Mirror Symmetry
}\vglue .75cm
  \centerline{and }
 \vglue .75cm
 \centerline{ Closed String Tachyon Condensation}
 }}
 \medskip

\centerline{Cumrun Vafa}
\vskip .5cm

\centerline{ Jefferson Physical Laboratory, Harvard University,
Cambridge, MA 02138, USA}
\medskip
\bigskip

\vskip .1in\centerline{\bf Abstract}

We study closed string tachyon condensation using the RG flow
of the worldsheet theory.  In many cases the worldsheet theory enjoys
$N=2$ supersymmetry, which provides  analytic control over the flow,
due to non-renormalization theorems.  Moreover, Mirror symmetry
sheds light on the RG flow in such cases.
  We discuss the relevant tachyon
condensation in the context of both compact and non-compact
situations which lead to very different conclusions.  Furthermore,  
the tachyon condensation leads to non-trivial 
dualities for non-supersymmetric
probe theories.

 \smallskip
\Date{November 2001}

\lref\beret{M. Bershadsky, A Johansen, T. Pantev, V. Sadov and C. Vafa,
`` F-theory, Geometric Engineering and N=1 Dualities,''
Nucl.Phys.{\bf B505} (1997)  153-164 [hep-th/9612052]; C. Vafa and
B. Zwiebach, ``N=1 Dualities of SO and USp Gauge Theories
and T-Duality of String Theory,'' Nucl.Phys. {\bf B506}
(1997) 143-156 [hep-th/9701015].}
\lref\ov{H. Ooguri and C. Vafa,``Geometry of N=1 Dualities in Four
Dimensions,'' Nucl.Phys. {\bf B500} (1997) 62-74 [hep-th/9702180].}

\def\fund{  \> {\vcenter  {\vbox
              {\hrule height.6pt
               \hbox {\vrule width.6pt  height5pt
                      \kern5pt
                      \vrule width.6pt  height5pt }
               \hrule height.6pt}
                         }
                   } \>
           }

\batchmode
  \font\bbbfont=msbm10
\errorstopmode
\newif\ifamsf\amsftrue
\ifx\bbbfont\nullfont
  \amsffalse
\fi
\ifamsf
\def\IR{\hbox{\bbbfont R}}

\def\IZ{\hbox{\bbbfont Z}}
\def\IF{\hbox{\bbbfont F}}
\def\IP{\hbox{\bbbfont P}}
\else
\def\IR{\relax{\rm I\kern-.18em R}}
\def\IZ{\relax\ifmmode\hbox{Z\kern-.4em Z}\else{Z\kern-.4em Z}\fi}
\def\IF{\relax{\rm I\kern-.18em F}}
\def\IP{\relax{\rm I\kern-.18em P}}
\fi

\lref\hiv{K. Hori, A. Iqbal and C. Vafa, ``D-Branes And Mirror
Symmetry,'' [hep-th/0005247].}

\lref\klebwi{I. Klebanov and E. Witten, ``Superconformal Field Theory on
Threebranes at a Calabi-Yau Singularity,''
Nucl.Phys. {\bf B536} (1998) 199-218 [hep-th/9807080].}

\lref\lnv{A. Lawrence, N. Nekrasov and C. Vafa, ``On Conformal Theories in
Four Dimensions,'' Nucl.Phys. {\bf B533} (1998) 199-209 [hep-th/9803015].}

\lref\mn{J. Maldacena and C. Nunez,``Towards the large N limit of pure N=1
super Yang Mills,'' Phys.Rev.Lett. {\bf 86} (2001) 588-591 [hep-th/0008001].}

\lref\witli{E. Witten, ``Phases of $N=2$ Theories In Two Dimensions,''
 Nucl.Phys. {\bf B403} (1993) 159-222 [hep-th/9301042].}

\lref\horv{K. Hori and C. Vafa, ``Mirror Symmetry,'' [hep-th/0002222].}

\lref\cfiv{S. Cecotti, P. Fendley, K. Intriligator and C. Vafa, ``A New
Supersymmetric Index,'' Nucl.Phys. B386 (1992) 405-452 [hep-th/9204102].}

\lref\cecov{S. Cecotti and C. Vafa, ``On Classification of N=2
Supersymmetric Theories,'' Commun.Math.Phys. {\bf 158} (1993)
569-644 [hep-th/9211097].}

\lref\dm{M. Douglas and G. Moore, ``D-branes, Quivers, and ALE
Instantons,'' [hep-th/9603167].}

\lref\ks{S. Kachru, E. Silverstein,``4d Conformal Field Theories and
Strings on Orbifolds,'' Phys.Rev.Lett. {\bf 80} (1998) 4855-4858
[hep-th/9802183].}


\newsec{Introduction}
A deeper understanding of non-supersymmetric string 
dynamics
seems to be the most fundamental obstacle to overcome in connecting
string theory with real world. For non-supersymmetric backgrounds, 
by
varying
moduli, one typically ends up in a situation with tachyons.  Thus
the question gets related to the fate of the closed string tachyons
upon their condensation.

\lref\CostaNW{
M.~S.~Costa and M.~Gutperle,
JHEP {\bf 0103}, 027 (2001)
[arXiv:hep-th/0012072].
}

\lref\fluxb{M. Gutperle and A. Strominger,
``Fluxbranes in String Theory,'' JHEP {\bf 0106}, 035 (2001)
[hep-th/0104136].}

\lref\posi{A. Adams,
J. Polchinski and E. Silverstein, ``Don't Panic! Closed String
Tachyons in ALE Spacetimes,'' hep-th/0108075.}

Recently the question of closed string tachyons has been addressed
in a number of situations \refs{\CostaNW ,\fluxb ,\posi}.
The aim of this note is to show that despite the fact that the target
is non-supersymmetric, very often the worldsheet in the NSR formulation
{\it is} supersymmetric, and for a wide class of examples
admits $N=2$ worldsheet supersymmetry.  Thus one can bring powerful
techniques developed in the context of 2d QFT's with $N=2$ supersymmetry
to bear on the question of tachyon condensation in superstring theories
in non-supersymmetric backgrounds.  In particular the RG flow of the
$N=2$ theories, whose F-terms are protected
by non-renormalizations theorems, suggest what the fate of the 
closed string tachyons are in many cases.

Here we make an assumption about the nature of string field theory,
which has proven to be rather successful in the context of open string
theories \ref\bsft{E. Witten,
``On Background Independent Open String Field
Theory,'' Phys. Rev. {\bf D46} 5467 (1992)\semi
S. Shatashvili, ``Comment on the Background Independent Open
String Theory,'' Phys. Lett. {\bf B 311}, 83 (1993). }\ref\shg{J.A. Harvey,
D. Kutasov and E.J. Martinec, ``On the Relevance of Tachyons,''
hep-th/0003101\semi
A.A.
Gerasimov and S.L. Shatashvili, ``On Exact Tachyon
Potential in Open String Field Theory,'' JHEP {\bf 0010}, 034 (2000)
[hep-th/0009148]\semi
D. Kutasov, M. Marino and G. Moore, ``Some Exact Results on Tachyon
Condensation in String Field Theory,'' JHEP {\bf 0010}, {045} (2000)
[hep-th/0009148].}.  Namely we view the relevant space for
closed string field theory to be the space of 2d QFTs and that the RG flow
in such spaces as indicative of dynamics of strings.  In this context we
identify the relevant 2d QFT's and deformations corresponding to tachyon
condensation and assume that there is an evolution in {\it physical time}
which can be identified with the {\it RG time on the worldsheet}.
This is of course consistent with the fact that fixed points of RG flow
are stationary solutions for classical strings.

The organization of this paper is as follows:  In section 2 we review
the gauged linear sigma model construction \witli\ and its mirror
description \horv\
and point out its
relevance to tachyons of orbifold theories.  In section 3 we apply the
ideas of section 2 to non-compact orbifolds with tachyons.  We consider
examples with complex dimensions 1,2 and 3.  In section 4 we consider
compact orbifolds with tachyons and consider the dynamics of tachyon
condensation in such cases.  For illustrative purposes we consider
the case of complex dimension 1 (orbifolds of $T^2$) and complex
dimension 3 (orbifolds of compact CY 3-folds).  In all such compact
cases
the internal theory loses some degrees of freedom as one would expect from
c-theorem
of Zamolodchikov.  In section 5 we show how the
consideration of probes in such theories
would lead to non-trivial non-supersymmetric dualities.  We give
some examples of non-chiral 4d non-supersymmetric dualities which
follows from this picture.  In section 6 we suggest some directions
for future work.

\newsec{Linear Sigma Model for Non-Compact Targets and its Mirror}

Worldsheet sigma models with $N=2$ supersymmetry have a powerful
description in terms of gauged linear sigma models \witli .  Let us
consider such a theory with a single $U(1)$ with charged matter
fields $(X_0,X_1,X_2,..,X_r)$ with charges given by
$$Q=(-n,k_1,k_2,...,k_r)$$
We take $n,k_i$ to be positive.  The theory is asymptotically free
when $n< \sum_i k_i$, and flows to a conformal theory
when $n=\sum_i k_i$.
The FI term for the $U(1)$
is naturally complexified, by combining it with the $\theta$-angle
to form a complex parameter $t$. 
When the theory is not conformal, $t$ can be traded with the scale.
In particular the UV fixed point corresponds to $t\rightarrow \infty$
and the IR fixed point corresponds to $t\rightarrow -\infty$.
A closely related theory is where we consider $Q\rightarrow -Q$.
This theory is the same as the above, except with $t\rightarrow -t$
with the role of UV, IR exchanged.  In particular suppose $Q$ corresponds
to a theory which is not asymptotically free.  Then $-Q$ corresponds
to a theory which is asymptotically free.  In this context $t\rightarrow
-\infty$ of the original model is the UV fixed point.

The geometry of the linear sigma model with a given $t$ can be analyzed
by solving the D-term constraints
\eqn\dterm{-n|X_0|^2+\sum_i k_i |X_i|^2=t}
modulo the $U(1)$ action 
$$(X_0,...,X_r)\rightarrow (X_0 e^{-in \theta},...,X_r e^{ik_r \theta})$$
which is the Higgs branch of the GLSM.  To be precise, 
the real part of $t$ appears in the above equation.  Note that
if we consider the limit $t<<0$ the equation \dterm\ implies
that $X_0$ takes a large vev.  In such a case the $U(1)$ is spontaneously
broken to $Z_n$.  The fields $X_i$ correspond to massless fields which
transform according to ${\rm exp}(2\pi i k_i/n)$ under the unbroken
$Z_n$ gauge symmetry.  Thus
we have that in the $t\rightarrow -\infty$ limit the GLSM is equivalent to
the ${\bf C^r}/
{Z_n}$ orbifold
$$(X_1,...,X_r)\sim (\omega^{k_1} X_1,...,\omega^{k_r}X_r)$$
where $\omega ={\rm exp}(2\pi i/n)$.  On the other hand in the limit
$t\rightarrow \infty$ the equation \dterm\ requires that not
all $X_i$, with $i=1,...,r$ are zero.  The equation \dterm\ 
together with the $U(1)$ gauge symmetry implies that the Higgs
vacuum with $X_0=0$ corresponds to the weighted projective space
$WP_{k_1,...,k_r}$
$$(X_1,...,X_r)\sim (\lambda^{k_1}X_1,..., \lambda^{k_r}X_r)$$
with $\lambda\not= 0$ (the phase of $\lambda$ corresponds
to the $U(1)$ action and its magnitude is related to the choice
of $t$).  Note that in this limit $t$ plays the role of the size (kahler
class) for $WP_{k_1,...,k_r}$.  The $X_0$ direction corresponds
geometrically to a non-compact bundle over this space which
is denoted by $O(-n)$.  Thus the target space is identified with the
total space of this bundle over the weighted 
projective space.  In other
words
the total space can be viewed in the limit of large $t>>0$ as the $r$
dimensional complex space given by
$$(X_0,X_1,...,X_r)\sim (\lambda^{-n}X_0,
\lambda^{k_1}X_1,...,\lambda^{k_r}X_r)$$
with $\lambda\not=0$ and not all $X_1,...,X_r$ are zero at the same time.

There is a convenient mirror description for this theory
which effectively sums up the gauge theory instantons \horv , roughly
by dualizing the phases of the fields $X_i$.  One obtains twisted
chiral fields $Y_i$ which are periodic variables 
$Y_i \sim Y_i+2\pi i$ and
are related to $X_i$ by
\eqn\real{|X_i|^2=Re Y_i}
and the theory becomes equivalent to a LG theory with
$$W=\sum_{i=0}^r {\rm exp}(-Y_i)=\sum_{i=0}^r y_i$$
(with $y_i=e^{-Y_i}$) subject to
$$y_0^{-n}\prod_{i=1}^r y_i^{k_i}=e^{-t}$$
(compare the absolute value of this equation with \dterm ).
We define $u_i$ by
\eqn\repl{u_i=y_i^{1/n}.}
From which we deduce 
\eqn\anre{y_0=e^{t/n}\prod_{i=1}^r u_i^{k_i}}
Note however that the change of variables \repl\
is well defined as long as we identify $u_i$ with a
$Z_n$ phase multiplication (since $Y_i$ are periodic).
However since the phase of $y_0$ is well defined, equation
\anre\ implies that the group we have to mod out by is
a subgroup of $(Z_n)^{r}$ preserving the monomial $\prod_{i=1}^r
u_i^{k_i}$, which is thus a group $G=(Z_n)^{r-1}$. Thus we have
found that in terms of $u_i$, after eliminating $y_0$
from the superpotential in terms of the $u_i$ the theory is
equivalent to
\eqn\finsu{W=[\sum_{i=1}^r u_i^n+e^{t/n}\prod u_i^{k_i}]//G}
where $G=(Z_n)^{r-1}$ is the maximal group preserving all the monomials.
This result was derived exactly as presented here in \horv .
Note that we have to be careful to consider $U_i$ as the natural
variables, where $u_i=e^{-U_i}$.  If $\sum k_i=n$ the
theory is conformal and that is reflected in the fact that the
above superpotential admits an R-symmetry in this case.  
 The RG flow corresponds to $W\rightarrow \Lambda^{-1}W$
where $\Lambda$ denotes the energy scale; this is due
to non-renormalization of F-terms which implies that the 
scaling is given by the naive classical scaling
given by the dimension $\int d^2x d^2\theta$.  By a field redefinition
\eqn\fred{u_i\rightarrow {\Lambda^{1/n}} u_i }
this gives a running for $t$ given by 
\eqn\run{t(\Lambda)=t +
(\sum_{i=1}^r k_i-n){\rm log} \Lambda}
  This  in particular implies that if
$\sum_{i=1}^r k_i$ is less than (greater than ) $n$
the UV fixed point
 is equivalent to $t\rightarrow -\infty$
($t\rightarrow +\infty$) and the IR fixed point 
 is equivalent to $t\rightarrow +\infty$
($t\rightarrow -\infty$).  

\subsec{The Orbifold Point}
As discussed before, for $t\rightarrow -\infty$ the theory
is equivalent to ${\bf C}^r/Z_n$, which is a conformal theory.
The mirror theory becomes equivalent in this limit to the LG theory with
superpotential
$$[W=\sum_{i=1}^r u_i^{n}]//G$$
where the only information about the $k_i$ is encoded in the action
of $G$ on the fields--it acts on each field by multiplication with
an $n$-th root of unity subject to preserving the
 chiral field $T=\prod_{i=1}^r u_i^{k_i}$.
The theory ${\bf C}^r/Z_n$ will have $n-1$ twisted sectors, each of
which gives rise, in its lowest state to a twist field, which is
also an $N=2$ chiral field \ref\qns{C. Vafa, ``Quantum 
Symmetries of String Vacua,'' Mod. Phys. Lett. {\bf A 4},
1615 (1989)\semi
C. Vafa, ``String Vacua and Orbifoldized
LG Models,'' Mod. Phys. Lett. {\bf A 4} (1989) 1169.}\
 (see also \ref\cecva{S. Cecotti and C. Vafa,
``Massive Orbifolds,'' Mod. Phys. Lett. {\bf A 7}
1715 (1992).}).  The first twist field is identified
on the mirror LG theory with $T=\prod_{i=1}^r u_i^{k_i}$ and that
generates the chiral fields associated with the other twisted sectors.
Namely in the $l$-th sector we get $T^l$ as the corresponding
twist field.  Since $u_i^n$ has to have $N=2$ charge $1$
the charge of $u_i$ is $1/n$ and that of the twist field $T$ is 
$$Q_{T}=\sum_{i=1}^r {k_i\over n}$$
which is the expected result for the charge of the twist field in the
first twisted sector.  Note that $N=2$ superconformal algebra implies
that the left and right dimension of $T$ is $h_{T}={1\over 2}
Q_{T}$.  A generic deformation by all twist fields is
given by
$$[W=\sum_{i=1}^n u_i^n + \sum_{l=1}^{n-1} t_l T^l]//G$$
for some complex parameter $t_l$ representing the strength
of the deformation for the ground state in the $l$-th twisted sector
and $T=\prod_{i=1}^r u_i^{k_i}$.  In the context of type II
superstrings GSO projection restricts the allowed $t_l$ as will be
discussed below, to be compatible with a $W\rightarrow -W$ symmetry.
 In particular
one needs a definition of an order 2 operator
$(-1)^{F_L}$.  For the action to be invariant, we need $W\rightarrow -W$
 because $\int d\theta_L d\theta_R W$ should be invariant
and $\theta_L$ is odd under $(-1)^{F_L}$.  
Also what one means by $T^l$ is 
$\prod u_i^{[lk_i]}$ where $0\leq [lk_i]<n$ and is
equal to $lk_i$ mod $n$, this is to make the relevant chiral
field to be the lowest dimension twist operator.

In the context of type II superstrings, the mirror LG model
we have presented changes IIA/IIB if the complex dimension
of the space is odd, and otherwise maps IIA and IIB back
to themselves.  This correlates with the number of times
T-duality has been applied.

Now we are ready to discuss some examples.

\newsec{Examples of Tachyon Condensation in Non-Compact Targets}

As an illustration of the models we will consider some non-compact
examples including the ones discussed in \ref\atd{A.
Dabholkar, ``Strings on a Cone and Black
Hole Entropy,'' Nucl. Phys. {\bf B 439}, 650,
1995 [hep-th/9408098]\semi
A. Dabholkar, ``Tachyon Condensation and Black Hole Entropy,''
[hep-th/0111004].}\posi .  We will consider
tachyons in 1,2 and 3 complex non-compact directions.

\subsec{Tachyon Condensations in ${C/Z_n}$}

Consider the linear sigma models with charges $(-n,k)$.  
 As $t\rightarrow -\infty$
this theory is given by the orbifold of 
$C/Z_n$ given by $x\rightarrow e^{2\pi ik/n}x$.

In this case the mirror description \finsu\ is given by the LG model 
with superpotential
$$W=u^n+e^{t/n}u^k.$$
Note that here $G$ is trivial. We take $k<n$ and for simplicity restrict
attention to the case where $k,n$ are relatively prime.
 As discussed above,
the deformation by 
$u^k$ is equivalent to condensation of tachyon in the 1st twisted sector.
Note that this orbifold $C/Z_n$ is equivalent to $x\rightarrow
e^{2\pi i/n}x$ since $k,n$ are relatively prime,
which would be given by a linear sigma model with charges
$(-n,1)$\foot{The fact that $W=u^n$ is the mirror
of $C/Z_n$ can be obtained also from taking the large $k$ limit
of Kazama-Suzuki $SL(2)/U(1)$ models, as noted in \ref\horkap{
K. Hori and A. Kapustin, ``Duality of the
Fermionic 2d Black Hole and $N=2$ Liouville Theory as
Mirror Symmetry,'' JHEP {\bf 0108}, 045 (2001).}.}.  However what is the
1st twisted sector in the $(-n,k)$ linear sigma model corresponds
to the $k$-th twisted sector of the $(-n,1)$ model.  Thus from the
perspective of the $(-n,1)$ theory, the FI-term deformation
of the $(-n,k)$ theory is equivalent to tachyon condensation in the
  k-th
twisted sector.  The infrared flow of this theory, as follows
from \run\ takes us
to the theory with $W=u^k$ which is equivalent to $C/Z_k$.  This is
exactly the pattern of flow discovered in \posi . In the case
considered in \posi\ avoidance of tachyon in the bulk restricts
 $n$ to be odd.  
    The beautiful arguments in \posi\
were based on the analysis in two regimes:  near the orbifold point, by
studying the deformation D-brane probes feel, and in the IR where the
target gravity regime was relevant.  Here we have found a framework
which interpolates between the two regimes in a way that we have
analytic control over the flow.  For example we can now recover the
RG evolution of the shell picture on the complex plane of the
 changing of the deficit
angle discussed in \posi .  Namely the space is given by
$$-n|X_0|^2+k |X_1|^2=t+(k-n){\rm log} \Lambda$$
modulo the $U(1)$ action,
where we have included the running of $t$ given by \run .  So in the
UV, where $\Lambda \rightarrow \infty$, we have a $C/Z_n$ geometry
and in the IR, where $\Lambda \rightarrow 0$ we have a $C/Z_k$ geometry.
  This is the same shell picture proposed in \posi .

In the context of type II superstrings, the existence of left/right
independent GSO projection further restricts the allowed $k$.  In particular
one needs a definition of an order 2 operator
$(-1)^{F_L}$.  For the action to be invariant, we need
an involution symmetry under which $W\rightarrow -W$. 
 This symmetry acts on $u\rightarrow
-u$ takes $W\rightarrow -W$ when $n$ and $k$ are both odd.
In the
context of type II strings this mirror LG description exchanges
type IIA strings with type IIB strings.
 We can
also translate this geometric description to the variables 
of mirror LG theory  using \real\ and \repl .

We can also consider deformations by all tachyon operators
from various twisted sectors at once.  This would correspond to
LG theory deformation with superpotential
$$W=u^n+\sum_{i=1}^{n-1}t_i u^i$$
where $t_i$ is the deformation in the direction of the 
tachyon coming from the $i$-th twisted
sector.  Here we are using mirror symmetry to connect
different linear sigma models for different values of $k$ 
into deformations of a single LG theory.   In the context
of type II superstrings, as discussed above, we restrict to the
case where $n$ is odd and only odd sector tachyons are turned on
(so that $u\rightarrow -u$ sends $W\rightarrow -W$).

\subsec{Tachyon Condensation in $C^2/Z_n$}
Next we consider the linear sigma model with
charge given by
$$(-n,k_1,k_2)$$ 
which in the $t\rightarrow -\infty$ goes over to
the $C^2/Z_n$ orbifold generated by
$$(X_1,X_2)\rightarrow (\omega^{k_1} X_1, \omega^{k_2} X_2)$$
where $\omega ={\rm exp} (2\pi i/n)$. 
The mirror theory is given by \finsu :
\eqn\deal{[W=u_1^n+u_2^n +e^{t/n} u_1^{k_1}u_2^{k_2}]//Z_n}
where $Z_n$ acts as n-th roots of unity on $u_i$ preserving
$T=u_1^{k_1} u_2^{k_2}$.  In the orbifold limit the LG superpotential
is given by $[W=u_1^n+u_2^n]//Z_n$ and the only input about $(k_1,k_2)$
comes from the action of $Z_n$ which preserves $T=u_1^{k_1}u_2^{k_2}$.
For the case $(k_1,k_2)=(1,1) $ 
or $(n-1,1)$  the theory is equivalent
to the target supersymmetric $A_{n-1}$ singularity, whose
twist field should give rise to a conformal theory.
However the above superpotential \deal\ seems to be
conformal (i.e. admit an $R$ symmetry) only for $(k_1,k_2)
=(n-1,1)$.  The point is that in the $Z_n$ orbifold leading
to spacetime supersymmetric
$A_{n-1}$, in the twisted NS sector there is a
tachyon
which is projected out by the GSO projection.  The difference
between the two cases here is exactly the choice of the GSO
projection in the twisted sector.
We can put this also directly in the language
of worldsheet conformal theory:  From the perspective
of the conformal theory we should see a 
tachyonic deformation, even for
the $(n-1,1)$ theory.  How is that realized
in the above LG theory?  This is realized by noting that
there is  a field ${\overline u_1} u_2$ which is invariant
under the $Z_n$ action for the $(n-1,1)$ theory, but it is not
chiral in the canonical realization of the $N=2$ algebra. 
 However in the orbifold limit
we have two decoupled $N=2$ systems and redefining the $U(1)$
charge of the first theory, gives an $N=2$ theory for which
${\overline u_1}u_2$ is a chiral field.  This would then be equivalent
to the deformations by $u_1u_2$ in the $(1,1)$ theory.

Just as in the $C/Z_n$ example we can consider
tachyon condensation in various twisted sectors and that
corresponds to further addition of the operator $T^l$ to
the superpotential.  Or, equivalently we can think of this
as the first twisted sector of the GLSM given by charges
$(-n,lk_1,lk_2)$, whose twist field corresponds to the FI
term deformation (or RG flow).  Let us consider first the simplest
case with $(k_1,k_2)=(1,1)$.  In this case we have charges given by
$(-n,1,1)$ and from the GLSM we know that in the IR the FI-term
flows \run\ so that $t\rightarrow \infty$.  The target space geometry
in this case will be the total bundle of $O(-n)$ over $P^1$, as discussed
before ($WP_{1,1}=P^1$).  Also the $P^1$ volume is given 
by $t$ and in the IR it becomes infinitely large.  Thus we obtain
$C^2$ as the end point of tachyon condensation in this case.

For a more complicated example let us consider general
$(k_1,k_2)$ with $k_1,k_2$ relatively prime
(the case $(1,k)$ is the case discussed in \posi\ with some shift $k\rightarrow
n-k$ in notation).

In this case the GLSM analysis suggests that in the IR the geometry
is given by the total space of an $O(-n)$ bundle over the
weighted projective space $WP_{k_1,k_2}$.  This is the space which
can be viewed as the complex space
$$(X_0,X_1,X_2)\sim (\lambda^{-n}X_0,\lambda^{k_1}X_1, \lambda^{k_2}X_2)$$
where $\lambda \not =0$, and $X_1,X_2$ cannot both be zero.
The compact part corresponds to the $X_1,X_2$ directions at $X_0=0$,
and the IR limit corresponds to making it infinitely large.  Except that
there are some orbifold points.  In particular at $X_1=X_0=0$ 
the geometry is locally
${\bf C^2}/Z_{k_2}$, as can be seen by considering $\lambda =e^{2\pi i/k_2}$.
The orbifold action is given by
$$(X_0,X_1)\rightarrow ({\rm exp}(-2\pi i n/k_2) X_0, {\rm exp}(2\pi i
k_1/k_2)X_1)$$
Similarly at $X_2=X_0=0$ we have another orbifold point
which locally is given by $C^2/Z_{k_1}$ given by 
$$(X_0,X_2)\rightarrow ({\rm exp}(-2\pi i n/k_1) X_0, {\rm exp}(2\pi i
k_2/k_1)X_2)$$
Thus the endpoint of the Tachyon condensation is given by a space
which has two orbifold singularities given by the two orbifold actions
above.  Moreover these two points are infinitely far away in space as the
volume of the $X_1,X_2$ space is going to infinty (and roughly speaking
they correspond to the north and south poles of a sphere).
Considering tachyon condensation in the $l$-th twisted
sector corresponds to replacing $k_1,k_2$ above by $lk_1,lk_2$. 
For generic flows this would end in
 non-supersymmetric orbifold fixed
points,
but for special choices it would have a supersymmetric endpoint
(for example if $n- k_1=0 $ mod $k_2$ or if $n- k_2=0$ mod $k_1$).  Of
course
at the new fixed points, in case they are
non-supersymmetric and include tachyons, we can also
use another tachyon direction to flow.  Since
 the order of the orbifold group  has gone down, i.e. $k_i<n$, this
procedure, i.e. continuing to condense tachyons if the corresponding
points are non-supersymmetric,
 would iteratively end up at a supersymmetric orbifold point or free space.

The geometry of the IR fixed point above
 can also be seen in the mirror LG description.  We have 
$$[W=u_1^n+u_2^n +e^{t/n}u_1^{k_1}u_2^{k_2}]//Z_n$$
In the IR limit the last term dominates, but this does not uniquely
fix the IR R-charge of the fields. In order to do that we have to go
to regions in field space where either $u_1$ is small or $u_2$ is small,
which would give limiting LG by dropping one of the first two terms.
This is the mirror of the two orbifold points we found above.  In particular
if we consider the regime $u_2\sim 0$ and
$u_1^n \sim
e^{t/n}u_1^{k_1}u_2^{k_2}$  (which can be arranged since $t>>0$)
then
 we have
$$[W\sim u_1^n+e^{t/n}u_1^{k_1}u_2^{k_2}]//Z_n$$
It is convenient to define 
$$v_1=u_1^{n/k_2}$$
$$v_2=e^{t/nk_2}u_1^{k_1/k_2}u_2.$$
The single valuedness of $v_i$ induces the $Z_n$ action above.
However the single valuedness of $u_1^n$ and $u_1^{k_1} u_2^{k_2}$ implies
that $v_1,v_2$ are obrifolded by $Z_{k_2}$ 
while preserving $v_1^{k_1}v_2^{-n}$, i.e., in this limit
we have
$$[W=v_1^{k_2}+v_2^{k_2}]//Z_{k_2}$$
where $Z_{k_2}$ preserves $v_1^{k_1}v_2^{-n}$.  This is the mirror of
$C^2/Z_{k_2}$ with action $(X_1,X_0)\rightarrow ({\rm exp}^{2\pi i k_1/k_2} X_1,
{\rm exp}^{-2\pi i n/k_2}X_0)$ obtained above near $X_1\sim X_0 \sim 0$.
Similarly by analyzing the limit where $u_1\sim 0$ and $u_2^n \sim
u_1^{k_1}u_2^{k_2}e^{t/n}$
we obtain the other orbifold point. 

Just as in the $C/Z_n$ analysis we can generalize the shell picture.
  This can also be stated in the original variables
of the linear sigma model, namely we consider the space of solutions
$$-n|X_0|^2+k_1 |X_1|^2+ k_2|X_2|^2=t+(k_1+k_2-n){\rm log} \Lambda$$
modulo the $U(1)$ action, where we have plugged in the RG flow of $t$
in accordance with \run .  

Considering the flow by the tachyon in the $l$-th twisted
sector, as noted before replaces $(-n,k_1,k_2)\rightarrow (-n,lk_1,lk_2)$
to which the above analysis applies.  We can also consider a linear
combination of fields in the twisted sector, for which the mirror
description is the most suitable one and given by a LG theory with
superpotential
$$[W=u_1^n+u_2^n+\sum_l t_l u_1^{lk_1}u_2^{lk_2}]//Z_n$$
Note however, if $lk_i\geq n$ then the corresponding twist
field is not the lightest state in that tachyon sector.  To obtain the
lowest state, we should replace $lk_i\rightarrow [lk_i]$ where
$[lk_i]=lk_i $ mod $n$ and $0\leq [lk_i]<n$.  In the following
we shall not bother putting brackets around $lk_i$ but that is what
is implied.  Note that $u_1^{[lk_1]}u_2^{[lk_2]}$ is 
also $Z_n$ invariant because $u_1^{mn}$ and $u_2^{m'n}$ are $Z_n$
invariant for all integers $m$ and $m'$.

In embedding this worldsheet theory in 
type II superstrings, we should also 
consider GSO projection.  If $n$ is odd, then the $(-1)^{F_L}$
can be taken to act as $u_i\rightarrow -u_i$, and we can, without
loss of generality assume $k_1+k_2$ is odd (if not, as discussed
before we can change one of them to $k\rightarrow n-k$ and this would then
be satisfied).  Then $T=u_1^{k_1}u_2^{k_2}$ is also odd
under $(-1)^{F_L}$, and so is
all the odd powers $T^l$ for $l$ odd.  In this case the deformations
are restricted to odd $l$ above.  If $n$ is even, we can assume $k_i$
are not both even (otherwise we divide all three by factors of $2$).  If
$k_i$ are both odd, then we define the $(-1)^{F_L}$ to act
as $u_1\rightarrow e^{i\pi (n-k_2)/n} u_1$
and $u_2\rightarrow e^{i\pi k_1/n}u_2$; this is an order 2 operation
on the invariant fields which are generated by $u_i^n$ and $u_1^{k_1}
u_2^{k_2}$.  In this case, again we can deform by all odd powers
of $T$, compatible with $W\rightarrow -W$ symmetry.  
The case when one of the $k_i$'s is odd and
the other even, will lead to a tachyon in the bulk which we are excluding.
Note that the mirror LG theory we have obtained here is for the same
IIA or IIB superstrings (since we have applied T-duality twice).

\subsec{$C^3/Z_{n} $ Tachyon Condensation}
Clearly we can consider many other cases.  For example consider
in complex dimension 3 the GLSM with charges
$$(-n,1,1,1)$$
If $n=3$ this is conformal and corresponds to the $O(-3)$ geometry
over $P^2$ or its orbifold limit as $t\rightarrow -\infty$ given
by $C^3/Z_3$.  For $n>3 $ the theory corresponds to a non-conformal
theory.  It corresponds to $C^3/Z_n$ with a tachyon which condenses.
Again in the IR the theory has a growing $P^2$ with infinite volume.
Thus in the IR it flows to flat $C^3$. The mirror description is
given by
$$[W=u_1^n+u_2^n+u_3^n+e^{t/n}u_1u_2u_3]//(Z_n\times Z_n)$$
where the $Z_n$ acts as $n$-th roots of unity on each field
preserving all the monomials.  Similarly we can consider more general
charges $(-n,k_1,k_2,k_3)$ which similarly to the $C^2/Z_n$ case would
end up with flat $C^3$ modulo three points infinitely far away
each of which is an orbifold of the type $C^3/Z_{k_i}$.  For example
for the $C^3/Z_{k_1}$ action it is given by phase action
on $X_0,X_2,X_3$ given by
$$(X_0,X_2,X_3)\rightarrow (\omega^{-n}X_0,\omega^{k_2} X_2,\omega^{k_3}X_3)$$
with $\omega^{k_1}=1$.  Again we can trace in the mirror language
where the corresponding regions come from, as in the $C^2/Z_n$ case.

In the context of superstrings we can again consider the condition
of being able to define $(-1)^{F_L}$, as in complex dimensions 1 and
2
discussed above.  We leave this as an exercise to the reader.

\newsec{Compact Tachyons}
In the previous section we have discussed tachyon condensation
in the context of non-compact theories, and this has given us
another conformal theory with the same central charge.  This can
also be seen from the relation between the central charge of LG theory
with the charges of the chiral fields \ref\vw{C. Vafa
and N.P. Warner, ``Catastrophes and the Classification
of Conformal Theories,'' Phys. Lett. {\bf B 218} 51 (1989)\semi
W. Lerche, C. Vafa and N.P. Warner, ``Chiral Rings in $N=2$
Superconformal Theories,'' Nucl. Phys. {\bf B324}, 427 (1989). 
}\ref\mart{
E. Martinec, ``Algebraic Geometry and
Effective Lagrangians,'' Phys. Lett. {\bf B 217} (1989) 431.}
$\hat c=\sum_i (1-2 q_i)$.
In the case at hand the relevant fields $Y_i$ have zero charge
(more precisely they have a ``logarithmic charge''
because they appear as $e^{-Y_i}$ in the superpotential
which has charge 1) so the
naive counting of the fields gives the complex dimension of the theory\foot{
If one considers the GLSM with charges $(-n,1)$ and add a superpotential
$X_0 X_1^n$ to the original sigma model, then according to the arguments
of \horv\ the mirror superpotential will still be given in the UV by $W=u^n$ but now
the good variable is $u$.  This would be the $n$-th minimal model of
$N=2$ SCFT \refs{\vw , \mart}.}. 
In particular the IR flow discussed above does not change the central
charge of the theory.  This was noted in \posi\ where it was emphasized
that the non-compactness allows one to evade the Zamolochikov's
c-theorem \ref\anpo{J. Polchinski, ``Scale and Conformal
Invariance In Quantum Field Theory,'' Nucl. Phys.
{\bf B 303}, 226 (1988).}.  In the compact case, as noted in
\posi\ one would expect
the IR flow to have a smaller central charge, and in the generic
case to become a purely massive theory.  Here we will show
that this is indeed
the case.  Furthermore 
we can also address the RG flow for such cases
using
mirror
symmetry techniques.

Let us consider the compact examples of dimensions 1 and 3.

\subsec{$T^2/Z_n$ Tachyon condensation}
Consider conforaml theory for $T^2/Z_n$
for $n=3,4,6$.  
The complex moduli of these tori are frozen, due to the
existence of the symmetry to be $\tau =e^{2\pi i/3},i,e^{2\pi i/3}$
respectively.  The Kahler moduli is free to vary.  
These theories will have tachyon fields in the
twisted sector and we can condense
them.  We are interested to study the RG flow
of the worldsheet theory upon such tachyon condensations.

Let us first discuss $T^2/Z_3$.  This is mirror to N=2, LG theory
with
$$W=x^3+y^3+z^3+axyz$$
where  $a$ is mirror to the Kahler structure of $T^2/Z_3$ and there is no
analog of the Kahler structure on the mirror because complex structure
is frozen.  To see this note that the LG/CY correspondence
\ref\gvw{B. Greene, C. Vafa and N.P. Warner,
``Calabi-Yau Manifolds and Renormalization Group Flows,''
Nucl. Phys. {\bf B324}, 371 (1989).}\ref\maral{E. Martinec,
``Criticality, Catastrophes and Compactifications,''
V.G. Knizhnik memorial vol. (L. Brink, et al. eds.) 1990.}\ 
 states that
$$[W=x^3+y^3+z^3+axyz]/Z_3 =T^2$$
where the complex structure of the $T^2$ is related to $a$ and Kahler
structure is fixed to be at the point with the $Z_3$ symmetry \qns .
Or, using the Kahler/Complex structure exchange of $T^2$ this implies
that this is equivalent to $T^2$ with complex structure frozen at the
$Z_3$ symmetric point and where $a$ is related to the Kahler parameter.
This is the way we wish to view the above LG theory. 
  Now as usual one can mod out by quantum symmetry on both sides
\qns .
Quantum symmetry on left gives back $x^3+y^3+z^3+axyz$.  On the right (i.e.
the
mirror) the quantum symmetry is a classical symmetry (this is due to the
usual
winding/momentum exchange) and so we obtain $T^2/Z_3$ which is the proof of
the
statement made above.  Note that the fields $x,y,z$ in the LG  theory
map
to the three tachyon fields of $T^2/Z_3$ at the three
fixed points of the $Z_3$ action. Adding them to the $W$ we have
$$W=x^3+y^3+z^3+a xyz + (t_1 x+t_2 y+t_3 z)$$
which will flow in the IR to a massive theory. Thus the $T^2$ formally should
disappear,
unlike the non-compact case.  Note that here
$ x,y,z$ are the good variable, whereas
in the
non-compact case $Y_i$'s are the good variable, where $y_i=e^{(-Y_i)}$
appears in the superpotential.
In the context of type II superstrings, we can define left/right
independent GSO projection by having $(x,y,z)\rightarrow -(x,y,z)$
under $(-1)^{F_{L}}$, which is compatible with the above deformation.

Similarly the other two orbifolds of $T^2$ are mirror to
$$W=x^4+y^4+z^2+a xyz \qquad for \qquad T^2/Z_4$$
$$W=x^6+y^3+z^2+a xyz \qquad for \qquad T^2/Z_6$$
where the twist fields are identified with $x,y,z$. Note that
this is consistent with the geometric fact that $T^2/Z_4$ has two
$Z_4$ fixed points and one $Z_2$ fixed point, and $T^2/Z_6$ has
one $Z_6$, one $Z_3$ and one $Z_2$ fixed point.
 Just as above
one can consider tachyon condensation for these theories.  However
in embedding in type II strings, since we have even order modding out,
this will lead to tachyon in the bulk.

\subsec{Tachyon Condensation on orbifolds of CY 3-fold}
Consider the LG theory given by
$$W=x_1^5+x_2^5+x_3^5+x_4^5+x_5^5+\psi x_1x_2x_3x_4x_5$$
If we modded this out by a $Z_5$ acting simultaneously
on all 5 fields by multiplication with $e^{2\pi i/5}$, it would be mirror
to the
$(Z_5)^3$ orbifold of the quintic \ref\greenepl{M.R. Greene 
and B.R. Plesser, ``Duality in Calabi-Yau Moduli Space,''
 Nucl. Phys. {\bf B 338}, 15 (1990).}.
  However we
can undo
the extra $Z_5$ in here as in the $T^2$ case, which on the mirror
corresponds to taking a $(Z_5)^4$ orbifold of the quintic not preserving
the holomorphic 3-form.  Thus we have a mirror symmetry on quintic modded
out by $(Z_5)^4$ with the above LG model without any modding out.
This leads to tachyonic modes in the geometry which just
as in the $T^2$ case can be identified with the chiral fields $x_i$
of the above LG model.  Again generic deformations by them will give a purely
massive theory.  However, we can also obtain intermediate situations.
For example if we deform by GSO allowed term
$\sum _i x_i^3$ (i.e. by using
tachyons in the triply twisted sector) we obtain a conformal
theory in the IR with complex dimension ${\hat c}=5/3$.
This has a piece which can be identified with the mirror of $T^2/Z_3$.
Or if we deform further by $x_4+x_5$ it would be exactly a $T^2/Z_3$
theory in the IR.  So we will have eaten up two
complex dimensions in this flow.

\newsec{Consequences for non-supersymmetric Dualities on Brane Probes}

Brane probes, possibly wrapped over non-trivial cycles
of compactification geometry, have been a source of a great
deal of interaction between field theory results and string theory.
In particular in \refs{\beret, \ov }\
it was argued how studying wrapped branes for type IIA, B strings
on Calabi-Yau threefold leads to Seiberg duality.
The idea there is to study the field theory living on the brane
as Calabi-Yau moduli vary.  
  This has
been extended to more non-trivial geometries recently \ref\cfikv{F.
Cachazo, B. Fiol, K. Intriligator, S. Katz and C. Vafa,
``A Geometric Unification of Dualities,'' hep-th/0110028.} 
\ (see also the related work \ref\plb{C. Beasley and M. Plesser,``Toric Duality Is Seiberg Duality,''
[hep-th/0109053].}\ref\hanse{B. Feng, A. Hanany, Y. He, A. Uranga, ``Toric Duality as
Seiberg Duality and Brane Diamonds,'' [hep-th/0109063].}).  

One can also follow the same idea in this context, namely
we consider probes in non-supersymmetric theories and ask what
we can learn about them from the analysis we did, about the fate
of the tachyon condensation.  We make an assumption here, which
we find plausible, but cannot justify. 
 We assume that {\it the IR dynamics of the
probe theory is the same as the theory on the probe 
after tachyon condensation in the target corresponding to the IR
limit of the worldsheet theory.}

One can provide many examples of such dualities.  To illustrate
the point, 
however,
we will limit ourselves to a single class of examples which we
develop in detail.
Consider the orbifold $C^3/Z_n$ which we studied in section
3, namely the one corresponding to GLSM with charges
$(-n,1,1,1)$.  As $t\rightarrow -\infty$ this is the orbifold
theory and the quiver theory on it can be deduced as discussed
in the general context in \ref\dougm{ M.~Douglas and G.~Moore,
``D-branes, Quivers, and ALE Instantons,''
hep-th/9603167.}\ and in the
present case in \ref\kas{S. Kachru, E. Silverstein,``4d Conformal Field Theories and
Strings on Orbifolds,'' Phys.Rev.Lett. {\bf 80} (1998) 4855-4858
[hep-th/9802183].}\ref\lnv{A. Lawrence, N. Nekrasov and C. Vafa, ``On Conformal Theories in
Four Dimensions,'' Nucl.Phys. {\bf B533} (1998) 199-209 [hep-th/9803015].}.
We will assume that $n=3k$ with $k$ odd.
The charges are given by
$(-3k,1,1,1)$.  This can also by viewed as a $Z_k$ orbifold
of the supersymmetric $C^3/Z_3$ orbifold.

Let us first consider the supersymmetric case $(-3,1,1,1)$.
If we consider $N$ D3 branes on this singularity, this
gives an ${\cal N}=1$ theory studied in \kas\ with $U(N)^3$ gauge
symmetry with three chiral matter fields $A^i_{12}$
$A^i_{23}$ and $A^i_{31}$, where $1,2,3 $ denote the three
gauge groups and the ordered subscripts on the fields represent
which bifundamental representation they transform as, and $i$
runs over three flavor values. There is in addition a superpotential
given by
$$W=\epsilon_{ijk} {\rm Tr} A^i_{12}A^j_{23}A^k_{31}.$$
As we go away from the orbifold point, i.e. from $t\rightarrow -\infty$
towards positive $t$, as it passes through 0, 
the probe theory
changes to its Seiberg dual as argued in \cfikv.  In particular
suppose node $1$ corresponds to the cycle which shrinks.  Then the dual
probe theory will again be ${\cal N}=1$ and will have gauge group 
given by $U(2N)_1\times U(N)_2\times U(N)_3$, and with matter fields 
$B^i_{21}, B^i_{13}, B^{ij}_{32}$, where $i,j=1,2,3$ are flavor indices
and $B^{ij}_{32}$ is symmetric in its $ij$ indices (i.e. there
are 6 such fields).  The superpotential for this dual theory is given
by
$$W=Tr B^{ij}_{32}B^i_{21}B^j_{13}$$
This corresponds to dualizing the first gauge group.

Now we consider the non-supersymmetric case corresponding
to $(-3k,1,1,1)$ (we take $k$ to be odd to avoid
tachyons in the bulk).  The probe theory on the orbifold point
for $N$ D3 branes is easy
to write down, and it will correspond to $U(N)^{3k}$ gauge group
with matter fields and interactions that can be written down
easily.
 This theory preserves no supersymmetry.
Under the RG flow, as discussed in section 3, the bulk
is expected to pass through $t=0$ after which case we would
expect to get a new theory on the probe.  This theory on the
probe should be a dual description of the non-supersymmetric theory.
The main question is how to find what this probe theory is {\it after}
this non-trivial transition.

In order to answer this question, recall that
we can also view the bulk theory as 
  the $Z_k$ orbifold of the supersymmetric $C/Z_3$
as noted before.  At the level of the brane
the $Z_k$ symmetry acts by 
$$g=e^{2\pi i (3R)/k}$$
where $R$ represents the $R$ charge, which for all the
$A^i$ fields is $2/3$.    We can also obtain the quiver theory
at the orbifold point
by acting on the $(-3,1,1,1)$ quiver theory by $Z_k$
by the usual rules discussed in \dougm , which
of course yields the same answer as we would obtain
by considering $C^3/Z_{3k}$.  But now, we also have a natural
proposal for what the dual non-supersymmetric quiver theory is:
If we start with the duality in the supersymmetric case for $t<0$
going to $t>0$, and if we mod out on both sides by $Z_k$, we may
expect the duality to continue in the non-supersymmetric case.
Examples of this have been seen in the context of 
large $N$ duals of non-supersymmetric
quivers (including the one under discussion) where the
AdS/CFT duality commutes with non-supersymmetric orbifolds.
In particular evidence in this direction was presented where
for example it was shown in \ref\bkv{M. Bershadsky,
Z. Kakushadze and C. Vafa,
``String Expansion as Large N Expansion of Gauge Theories,''
 Nucl.Phys. {\bf B 523}, 59 (1998)\semi
 M. Bershadsky and A. Johansen, ``Large N limit of orbifold field theories,''
 Nucl. Phys. {\bf B536} 141 (1998).}\ that the small
`t Hooft coupling analysis of large $N$ conformality, agrees with
the large `t Hooft coupling analysis of \kas\ on the AdS side.
One thus expects the same to be true here.

Note that the dual theory has $R$ charges of chiral fields given
by $R_{B^i_{13}}=R_{B^i_{21}}=1/3$, $R_{B^{ij}_{32}}=4/3$, and using
this we can find the $Z_k$ orbifold of this quiver,
which we identify with the dual non-supersymmetric quiver theory
on the non-supersymmetric probe.  It would be interesting
to check the validity of this non-supersymmetric duality.
More generally  one is tempted to use this
idea to generate many more non-supersymmetric dual quiver theories
by modding out a supersymmetric pair of dual quiver theories
with an $R$ symmetry on both sides.  In fact this idea has already
been considered in \ref\shm{M. Schmaltz, ``Duality of Non-supersymmetric
Large $N$ Gauge Theories,'' Phys. Rev. {\bf D59}
 (1999) 105018 [hep-th/9805218].}\ and successfully
tested at large $N$.

\newsec{Open Problems}

One could apply the ideas in this paper to a number of different
situations.  For example, one can apply it to the question of 
tachyon condensations in the context of fluxbrane.  This
seems to yield a nice picture \ref\shg{J. David,
M. Gutperle, M. Headrick and S. Minwalla,
to appear.}.  One can also consider more elaborate geometries
than considered in this paper,
and study the corresponding dualities for non-supersymmetric
theories on the probe.

So far we have mainly concentrated on aspects of worldsheet theory
and concentrated on the RG flow of the worldsheet and
assumed that some target string dynamics represents this flow.
However, even within the framework we have studied
it is natural to ask if there is a full conformal theory
representing the internal RG flow.  In particular
can we write down a theory with a fixed central charge
as a fibered QFT where on the fiber the theory appears
not to be conformal but the totality of the theory
is conformal with the critical central charge
for string theory?  For example
one could look for solutions where the compact Calabi-Yau
in one region is smoothly connected to a region
where the CY has been eaten up! It is natural
to look for such conformal theories which preserve $N=2$
superconformal symmetry on the worldsheet.  In fact,
it turns out that it is possible to make a canonical
construction along these lines by promoting the complexified RG
parameter
to a chiral field.
For example, we consider
$$[W=\sum_{i=1}^r u_i^n+e^{X} \prod u_i^{k_i}]//G$$
in the non-compact case, or say
$$W=\sum_{i=1}^5 x_i^5 +e^{X} \sum_{i=1}^5 x_i$$
in the case of compact quintic orbifold.  We can view
$X$ as a non-compact extra complex space and
 one can assign
an R-charge to $e^X$ so that $W$ can flow to an
$N=2$ CFT.  This is basically the same as introducing the $N=2$
axion field to the system in the language of GLSM.
The central charge
is given by $\sum(1-2q_i)$ and is fixed.  This is not the temporal
dynamics one
usually
looks for in the Tachyon condensation, but nevertheless
it is a well defined conformal theory and a consistent
background for superstrings.  In such a
description
for $Re (X) <<0$ we are at one ``internal'' CFT and for
$Re (X)>>0$ we are at a very different one.  In a sense
this is a kind of ``Euclidean instanton'' corresponding
to tachyon condensation.

One may also expect the 2d QFT to be part of
  a consistent string field theory. This would be clearly
  rather desirable to develop.
   In such a framework we should have
in particular a notion of what a tachyon potential is.
In connection with what a right notion of tachyon potential
may be in the context of $N=2$ supersymmetric worldsheet
theories one has a natural thought which is currently under
study \ref\atva{A. Dabholkar and C. Vafa, work in progress.}.
The idea is to relate the axial charge of chiral
fields (which changes under the RG flow) to the tachyon potential.
The axial charge is precisely what gives the $-m^2$ of the tachyon
potential at the conformal points and so its flow should be
naturally related to the tachyon potential.  Luckily the axial
charge can be exactly computed in $N=2$ theories in terms
of $tt^*$ geometry developed in \ref\cevatt{S. Cecotti and 
C. Vafa, ``Topological Anti-topological Fusion,'' Nucl. Phys.
{\bf B324} (1989) 427.}. This is generally characterized
by solutions to an integrable system.  One would thus expect
to compute the tachyon potential in terms of solutions
to this integrable system.

\vskip 1cm

\centerline{\bf Acknowledgements}

I would like to thank E. Witten for many valuable
discussions.  I would also like to thank
M. Aganagic, N. Arkani-Hamed, A. Cohen, A. Dabholkar,
M. Gutperle, K. Hori, K. Intriligator,
S. Katz and S. Minwalla for comments on this work.

This research is supported in part by NSF grants PHY-9802709
and DMS-0074329.

\listrefs

\end